# Gaia: 3-dimensional census of the Milky Way Galaxy

Gerard Gilmore

Institute of Astronomy, University of Cambridge, Madingley Road, Cambridge CB3 0HA, UK.

Astrometry from space has unique advantages over ground-based observations: the all-sky coverage, relatively stable, and temperature and gravity invariant, operating environment delivers precision, accuracy and sample volume several orders of magnitude greater than ground-based results. Even more importantly, absolute astrometry is possible. The European Space Agency Cornerstone mission Gaia is delivering that promise. Gaia provides 5-D phase space measurements, 3 spatial coordinates and two space motions in the plane of the sky, for a representative sample of the Milky Way's stellar populations (over 2billion stars, being ~1% of the stars over 50% of the radius). Full 6-D phase space data is delivered from line-of-sight (radial) velocities for the 300million brightest stars. These data make substantial contributions to astrophysics and fundamental physics on scales from the Solar System to cosmology. Reliable parallax distances in astronomy were available for of order $10^4$ stars to milliarcsec (mas) precision in the 1980s, for of order $10^5$ stars to mas accuracy in the 2000s, and with Gaia for more than $10^9$ stars to 10μas accuracy. A knowledge revolution is underway.

1. **The Milky Way Galaxy- a Rosetta Stone for astrophysics**

Our Milky Way Galaxy is a large spiral galaxy, typical of those which dominate the light in the Universe. As all galaxies, it started its formation at very early times after the Big Bang, from initially low amplitude (sound wave) fluctuations in the smoothly-distributed then gravitationally dominant dark matter. These fluctuations grew under self-gravity, generating a hierarchical system of merging self-gravitating structures, each gravitationally bound by its dark matter but also containing a share of the baryons which later form the stars, chemical elements, planets and interstellar gaseous medium. This process on large scales is very well described by the standard ΛCDM cosmology, in which Λ represents the dominant (70% of the energy density today) dark energy, a not yet understood property of space-time which is accelerating space-time against gravitational attraction. CDM describes the cold dark matter (25% of the energy density today), an unknown (set of) slowly-moving (hence "cold") gravitating particles with extremely low cross-section to the standard three forces of nature (electromagnetic, weak and strong). The model also includes the remaining 5% of the energy density in baryons, which make up all that we can see and all that is described by modern physics. On the scales of galaxies and smaller (1Mega-parsec, Mpc, ~3.E+25m) baryonic physics becomes important, in particular the dependence of gas cooling on the abundances of the chemical elements, which are being created in supernovae, and effects on gas flows of energy inputs from supernovae and supermassive black holes, and much complex microphysics, generically labelled "feedback". In addition in a stellar disk galaxy, dynamical instabilities generate spiral and bar-like gravitational perturbations which can move stars far from their birthplaces. Unravelling the relative importances of this time- and place- dependent set of complex physical processes is a challenge for modern astrophysics. This is where our Milky Way can make a special contribution to our detailed knowledge: its neighbourhood the Local Group, and its constituents – satellite galaxies, younger thin and older thick stellar disks, a central bulge, a thin-disk bar and spiral arms, an ancient extended halo, current merger activity, a wide range of chemical elements heavier than helium created in a variety of stellar evolutionary processes, and the dominant dark matter – are the only parts of the Universe close enough for detailed census and analysis. And perhaps discovery of new physics, given that we do not understand 95% of what we measure.

**1.1** *the need for astrometry: distances in astronomy*

Astrophysics has been mapping the sky for millennia, so it is reasonable to ask what information is needed to address the top-level questions above, and why it is not yet available (Figure 1).

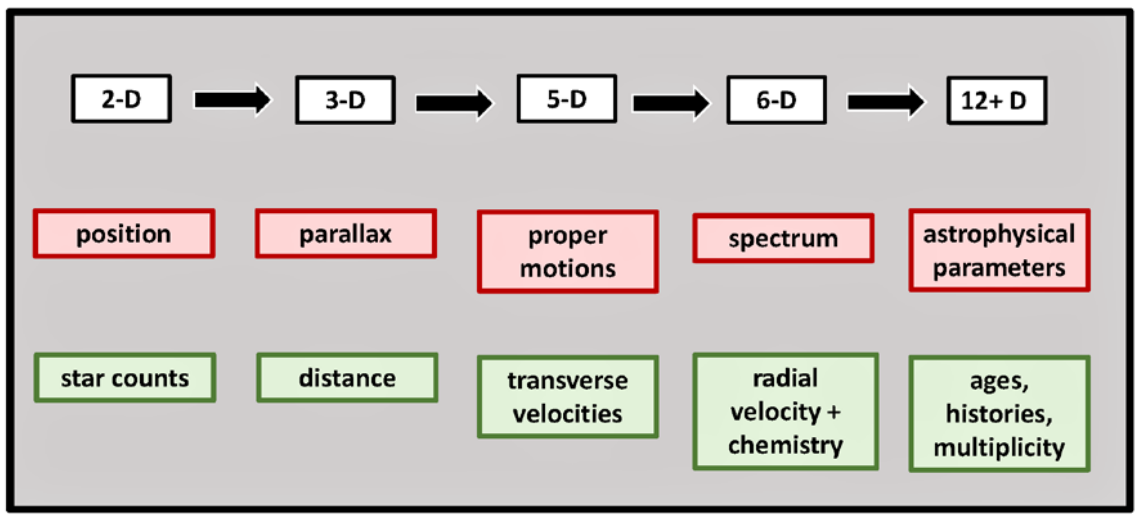

**Figure 1:** The sub-set of observable properties of an astrophysical source relevant to the Gaia mission. Each property is independent of the others and so may be thought of as dimensions.

The most basic information set is 2-D star counts, recording the number of stars as a function of position and brightness (the logarithmic magnitude scale was invented in Ptolemaic times for this reason). These counts show the Milky Way to be flattened, with the Sun close to the symmetry plane. More detailed analyses however require distances, which can be derived robustly only from parallax[1], the apparent annual motion of a star caused by movement of the observer on earth around the Sun during the year (Figure 2). This 3-D information allows determination of the 3-dimensional structure of the Galaxy. When these observations are repeated over time the intrinsic motion of the star ("proper motion", meaning the property of the specific star, cf "*propre*") additionally provides 2-D velocity data. Adding a line-of-sight "radial" velocity, which requires spectroscopic information, then provides 6-D phase space data for a star, with its 3-D configuration-space position and the 3 components of its space motion velocity vector. These measures can also determine multiplicity, either companion stars or a planetary system. 6-D data allow dynamics, quantifying the time-dependent kinematical structure of the Galaxy – spiral arms, dynamical bar – and through application of the coupled Poisson-Boltzmann equations, determinations of mass distributions, and discovery and mapping of cold dark matter. The spectroscopy required to determine a radial velocity can additionally provide information on the stellar properties and chemical elemental abundances. The chemical elements may be classified into a small number of families, defined by the process which created them – Big Bang (H, He), nuclear burning in stars: the p- r- s- and i- processes, and cosmic ray spallation. Thus chemistry essentially adds 4-5 more effective dimensions of information, which are related to the time and place of a star's formation, and the prior history of stellar elemental production, supernovae and gas flows in and out of the star formation region. Ideally one would like to add stellar age, but that is harder to determine.

[This discussion excludes most of the gas and molecular content of the Galaxy, better studied using radio and far-IR techniques, and also the hot universe, requiring UV and X-ray data. We restrict to optical studies here, largely of stars.] This is the information set potentially available to address the evolution of the Milky Way.

There is however an elephant in the room. Stellar distances are large, stellar parallaxes are correspondingly small, so much so that stars were routinely described as fixed stars, *stellae fixae*, through most of history. Although everything in the Universe moves (the Earth orbits the Sun at 30km/s, the Sun orbits the Milky Way at 230km/s), even precise measurement of proper motions is difficult, preventing large-scale implementation of our dimensional data set in Figure 1. The parallax distance to a star is $d(parsec) = pi^{-1}$, with pi in arcsec. Thus determining accurate distances to the centre of the Milky

---

[1] parallax defines the basic distance unit in astronomy, the parsec, that being the distance corresponding to a heliocentric parallax of one second of arc, equivalent to 206265 times the astronomical unit, the mean Sun-earth distance, or 3.0867E+16m, or 3.262 light years

Way, at 8kpc, implies parallax accuracy of order micro-arcseconds (µas), which is sub-picoradian. This can be achieved only by astrometry from space.

**1.2** *space astrometry*

The apparent path of a star over time is a convolution of the orbit of the star in the Galactic potential, the apparent parallactic motion of the moving observer, and often high order kinematic perturbations from a binary stellar companion and/or an exo-planet system (Figure 2).

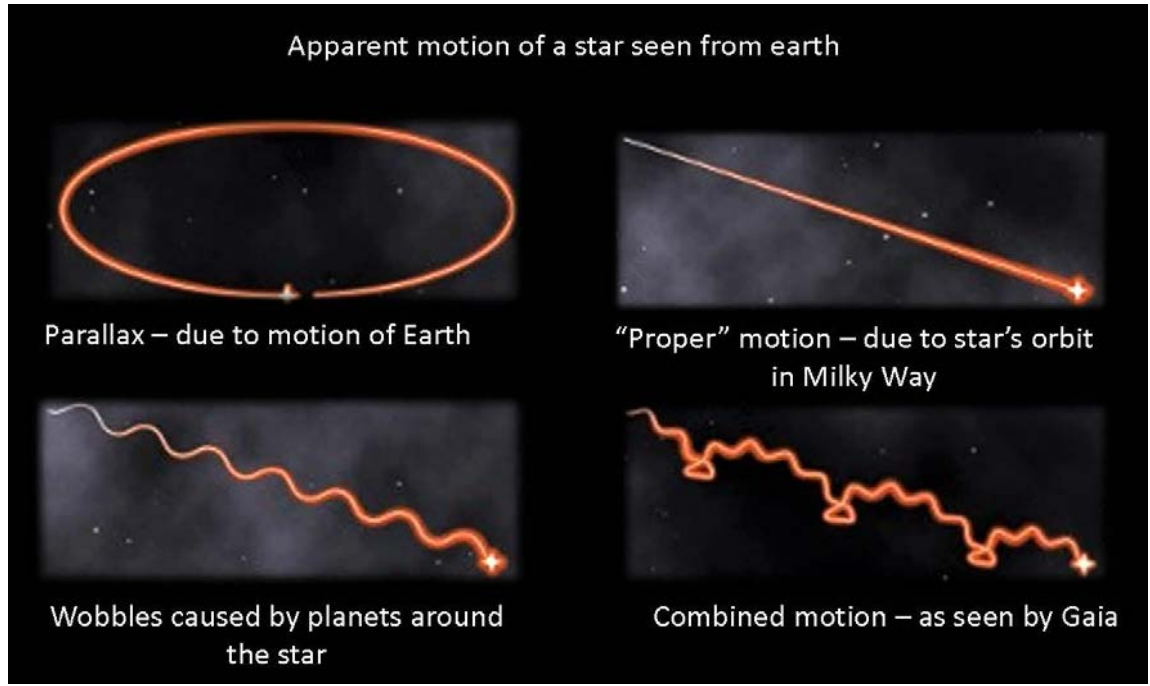

Figure 2. illustration of the several contributions to the apparent path of a star across the sky.

In order to model this path one requires at minimum six parameters (one per dimension in Fig 1 for a single star, being the 5 traditional astrometric parameters and a radial velocity), plus sufficient parameters to model multiplicity/planets, and clearly enough precise data over a sufficiently long time to allow a robust fit. The required duration of observations is defined by the annual parallax, the relevant (planetary) companion orbital periods, and the sensitivity limit required to measure suitably accurately the transverse proper motion of the star. There is an additional requirement, uniquely requiring space data – the parallax zero point.

The orientation of the parallactic ellipse in a line of sight depends on the sine of the viewing angle along a great circle to the Sun. That is essentially constant over a typical single telescopic field of view, so that one can measure only differential parallaxes along a single line of sight. Determination of a zero point must then be provided by comparison with an object at very large distance, typically a quasar or galaxy if available in the line of sight. Such sources are usually rare, and have however different image properties than those of stars, leading to poorly determined zero points. Provision of an absolute measurement requires simultaneous observation of two fields of view, separated by close to 90deg, with precise knowledge of the relevant "basic" separation angle. Having two telescopes at known angle allows local small-angle (differential) positional measurements to be connected to large angular separations. That is, one needs two telescopes feeding a single focal plane, in a gravity-invariant location, with no differential atmospheric refraction, *ie* in space. Proof of the viability of this approach was provided by the European Space Agency mission HIPPARCOS [1], launched in 1989 with catalogue publication in 1997, which delivered astrometry for $10^5$ bright nearby stars with milli-arcsecond (mas) accuracy and precision, and a photometric catalogue of 2.5e6 stars.

The success of HIPPARCOS led immediately to a proposal for a much more ambitious mission, which could address the top-level science ambitions listed above. This proposal built on technological advances, in precision spacecraft design and in high-efficiency large-format CCD devices: HIPPARCOS had a single-pixel detector, Gaia has a 938,000,000 pixel focal plane, the largest yet in space. Community support for the case allowed by extending to one billion stars, 10 microarcsec precision, and a sensitivity (magnitude limit fainter than V=20) sufficient to deliver a strong and diverse science case, led to adoption of the Gaia mission by ESA in 2000.

2. **Census of the Milky Way: the Gaia mission**

The Gaia spacecraft was designed and built by a consortium of largely European industries, led by Astrium (now known as Airbus Defence and Space). It was launched by a Soyuz-Fregat rocket (VS06) from the European Spaceport at Kourou, French Guiana on 19 December 2013. Following launcher separation the folded solar array was deployed, and the spacecraft left low-earth parking orbit to spend about one month in transit to a large Lissajous orbit around the L2 Lagrange point of the Sun-Earth system, some 150million km from Earth. Following in-orbit commissioning, full science operations began on July 25 2014. Full detailed descriptions of the Gaia mission are provided in the papers published with the Data Releases[2]. An up to date list of articles about the mission and those using Gaia data for science research is maintained as part of the ESA Gaia mission web site at https://www.cosmos.esa.int/web/gaia/publications.

2.1 *the name "Gaia"*

The name Gaia derives from one of the original 1993 mission proposals (Global Astrometric Interferometer for Astrophysics, GAIA; the other main proposal was called Roemer), although the spacecraft as built is not an interferometer. The initial acronym GAIA mutated into the name Gaia and survives. It turns out to be an appropriate name in its own right - motivating the logo on the fairing of the launch rocket (Fig 3) - after the ancient goddess as she appears, for example, in Hesiod's *Theogony 116/117 & 126/127* (Figure 3) who came into being after *Chaos* and generated the starry sky. One interpretation of her coming into being may be as a contrast to the unintelligible (*Chaos,* a gap, a wide opening) and as a generator of the explorable (the starry sky amongst many others). The early history of the Gaia mission is summarised in [3].

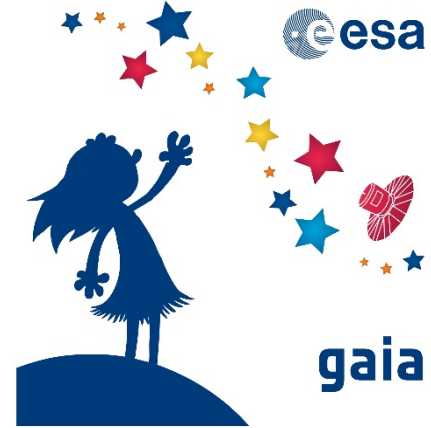

**Figure 3. LHS:** The description of Gaia from Hesiod's Theogony[2]. **RHS** The ESA Gaia logo from the launch-rocket fairing, the cover which protected Gaia during launch. Courtesy ESA.

## 2.2 *Gaia payload design*

Gaia's measurement principle relies on a two-telescope smoothly scanning spacecraft with a focal plane which allows precise location of the flux centroid and measurement of the crossing time of each target. These times, and associated image fluxes, are the one-dimensional along-scan stellar positions relative to the spacecraft axes. These (1-D) observation times are converted to astronomical coordinates by a simultaneous reconstruction of the spacecraft orientation as a function of time, and the instrument (payload) geometric calibration from focal plane through the opto-mechanical system to the sky. As all calibration and source astrometric parameters are derived from the same data, Gaia is self-calibrating. Conversion of focal-plane time to sky requires the payload system to behave smoothly and continuously during the measurement process, ideally stable at the few microarcsecond level on times of minutes to hours. For this reason Gaia is built of silicon carbide components on a super-stable sintered silicon carbide octagonal optical bench, with a 3 metre diameter sunshield isolating the cold payload from the service module warm electronics, and the spacecraft support and communications systems (Figure 4). On the principle that one

---

[2] "(116) In truth, first of all Chasm [*Chaos*] came to be, and then broad-breasted Earth [*Gaia*] (…)" - "(126) Earth [*Gaia*] first of all bore starry Sky, equal to herself (…)" [4]

designs for perfection and calibrates for performance, and that the basic angle between the two telescopes must be known at the microarcsecond level, Gaia includes what is perhaps the most precise laser interferometric metrology system yet on a spacecraft, the Basic Angle Monitor. The Basic Angle Monitor delivers differential measurements with accuracy 0.5µas each 15 minutes, corresponding to picometer displacements of the primary mirrors.

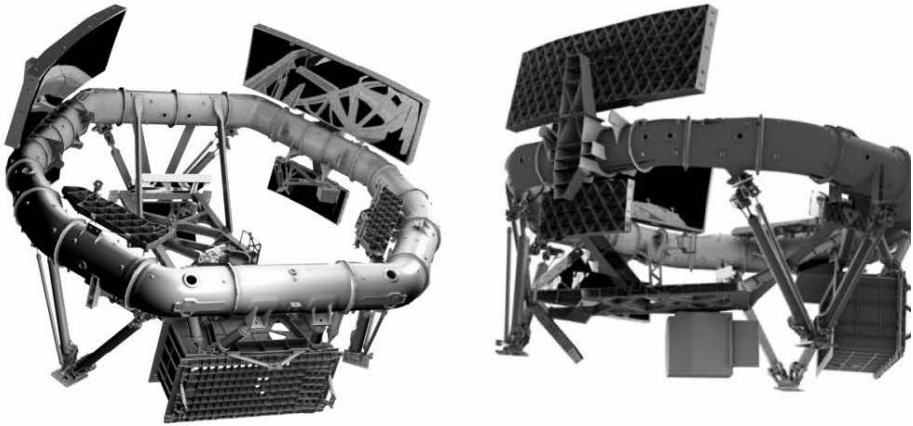

Figure 4: Schematic overview of the physical payload, without the protective cover. The octagonal ring is the optical bench supporting the structure. The rectangular primary mirrors of the two separate telescopes are mounted above this ring. Other mirrors are mounted below. The very large box below the ring holds the focal plane (cf Fig 5), its proximity electronics and a large cooling radiator. The smaller box visible in the RHS image in the middle below the structure is the (transmission) Radial Velocity Spectrometer. The sunshield is mounted below the structure shown here, with the service module below that. Most electronics boxes are located in the service module, which is located below the right-hand image above, and not shown here. Credit: ESA

2.2.1 *The two telescopes*

The two telescopes are identical three-mirror anastigmats, with 3 flat folding mirrors to contain a focal length of 35m, with primary mirror apertures of 1.45m by 0.50m, and with the two lines of sight separated by the basic angle of 106.5deg along the scanning circle. The two telescopes feed a single, very large, focal plane.

The rectangular mirrors are interesting, and have consequences in downstream data processing. As noted above, it is the along-scan time-flux data which constrains the astrometry. Across scan the spacecraft motion is controlled to generate just enough precession between adjacent great-circle scans to deliver full-sky coverage. The across-scan motions are less smooth and stable than those along scan, so they contribute negligibly to the astrometry. Given that, there is no need to collect the flux required for determination of a precise cross-scan position. Thus one can save mass and space by cutting off (ir)relevant parts of the primary mirrors, leaving a rectangle. The consequence is an elliptical point spread function, retaining full spatial resolution and photon-collection efficiency along the scan direction, and a scan-angle dependent spatial resolution.

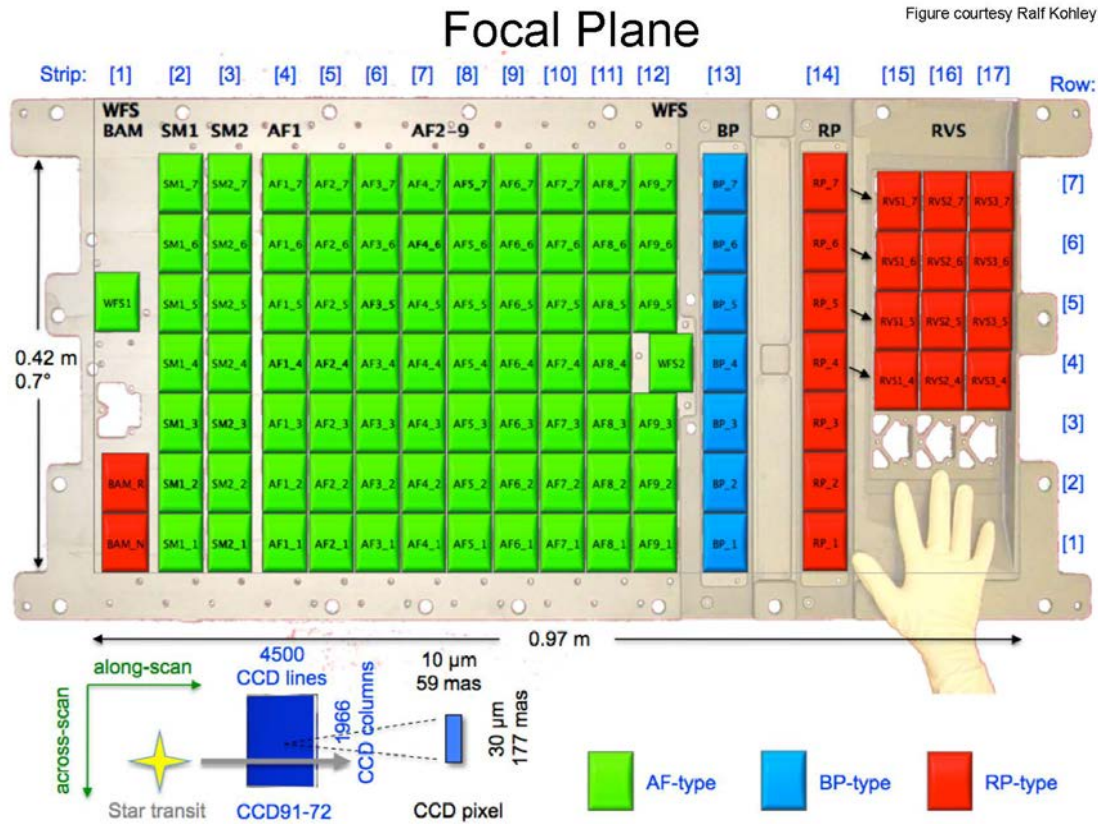

Figure 5. The Gaia focal plane assembly. Schematic view of the very large focal plane assembly, with background the actual Gaia support structure, and a cartoon hand for scale. The primary Gaia names for the activities are indicated. From the left labelled items include: the Wavefront Sensor (WFS) for the Basic Angle Monitor (BAM), the SkyMappers for each of Telescope one (SM1) and two (SM2), the nine rows of Astrometric Field (AF) CCDs, the Blue Photometer (BP), the Red Photometer (RP), and the three rows of CCDs for the Radial-Velocity Spectrometer (RVS). Figure courtesy of Airbus DS and Boostec Industries.

### 2.2.2  *The focal plane instruments*

The focal plane is a very special feature of Gaia, delivering five types of data while also being unusually large (Figure 5). The focal plane consists of 17 along-scan strips of detectors, with 7 across-scan rows. Each science-data detector is an e2v (now Teledyne e2v) CCD, variants of the CCD91-72 with 4500 lines and 1986 columns, with pixels being 10μm x 30μm (58.9mas along scan  x 176.8mas across scan) to match the primary mirror point spread function. In total the focal plane has 938,000,000 pixels. In the scan direction, the focal plane first delivers metrology: Shack-Hartmann wave front sensing for mirror focus and alignment, and the Basic Angle Monitor sensors. Then follow two columns of sky-mapper sensors, baffled to one column  per telescope. These image, detect, determine the flux and centroid, and classify all bona fide point-like sources – those with image full width at half maximum smaller than 0.7arcsec, with completeness limit about G=20.7, where "G" is the unfiltered native spectral response of the Gaia photometric system. For each detected source an electronic window of size matched to image brightness is established, with an associated integration time, which follows the image in TimeDelayedIntegration (TDI) mode across the focal plane.  Each image is then tracked across columns 4 to 12 (AF1, for Astrometric Field column one, to AF9), while the AF1 image is real-time analysed to confirm the detection results and check for moving sources (asteroids). Columns 13 and 14 have low-dispersion prisms mounted in front of them, blue (BP, for Blue Photometer: 330-680nm) and red (RP, for Red Photometer: 640nm-1050nm) to deliver low-resolution spectrophotometry, with optimised CCD sensitivies for each detector set. The final 3 CCD columns are of 4 rows, and are fed by a spectrograph, known as the Radial Velocity Spectrometer - RVS, delivering slitless R~11700 spectra over the wavelength range 845nm-872nm, but restricted to brighter stars.

The Gaia CCDs have dynamic range/full-well capacity around 190,000e. The system dynamical range is further extended by an ability to select effective integration times per CCD ranging from 4500 TDI lines, for typical faint sources and corresponding to an integration time of 4.42sec, down to 2 lines, for very bright sources. This allows an observational dynamic range of $5 <\sim G <\sim 20.7$. At the bright end, although images are saturated, sufficient information is being recorded so that astrometry and spectroscopy will be possible to significantly brighter limits. There is obvious scientific merit in ensuring that astrometry of the apparently brightest stars is on the same system as that for most stars so that astrometry for even brighter stars is desirable. Although the reduction system is not proven as yet, special imaging observations (known as SIF images) are being obtained for all bright stars, so that in principle Gaia will be complete at bright magnitudes, eventually delivering data over the dynamic range $-1<G<21$, a dynamic range approaching ten orders of magnitude. The astrometric field (white light, AF) integrations provide the main astrometric/photometric measurements. Spectrophotometry (BP/RP) is obtained to provide spectral energy distribution data which is needed for astrophysical analysis/source classification, and also to improve the primary astrometric drivers of the Gaia mission.

Although the mirrors are of high quality, and the optical system is achromatic, residual wavefront mirror polishing errors generate chromatic aberrations which affect the potential astrometric accuracy. Thus the source spectral energy distribution is necessary input during modelling of the astrometric solution. The prism dispersion varies from 3 to 27 nm/pixel in BP and 7 to 15 nm/pixel in RP. The dynamic range of the BP/RP spectrophotometry is limited at the bright end by saturation – though supplementary data for the few bright stars can be obtained independently of Gaia -  and at the faint end largely by crowding, as discussed below, when in any case the astrometry is very photon-starved. Spectra from the RVS are limited to brighter sources than for the photometry. For the few million stars brighter than G~12.5 astrophysical parameters and element abundances can be derived reliably from single-transit observations. At fainter magnitudes just radial velocities can be measured, by stacking individual very low signal-noise transits, for several hundred million stars to G~16. These data provide the key $6^{th}$ phase space dimension in Figure one, and also provide an essential $6^{th}$ astrometric parameter for modelling nearby fast-moving stars, where perspective acceleration over the Gaia observing duration is significant.

### 2.2.3 *Scanning the sky*

Gaia is in a Lissajous halo orbit of amplitude of order $10^5$km around the second Lagrange point (L2) of the Sun-Earth-Moon system, which is some 1.5 million km from Earth away from the Sun. As such it appears as a $20^{th}$ magnitude star close to the equator at local midnight from Earth. This orbit ensures that the Earth remains sufficiently aligned with the phased-array communications antenna, while delivering a relatively stable thermal environment. The orbit, with one planned correction in 2019, avoids eclipses. The sky-scanning law of Gaia is a key aspect of the astrometric performance, designed to optimise delivered final astrometric accuracy. Gaia adopts uniform revolving scanning, maximising its uniformity of sky coverage. The spin rate about the spacecraft spin axis is nominally 60 arcsec/sec (actual is 59.9505), matching image motion to CCD clocking speed. The solar-aspect angle is 45deg between the Sun – actually a nominal Sun position referred to the global astrometric reference frame so independent of Gaia's orbit - and the instrument spin axis, maximising parallax sensitivity. The along-scan parallax displacement is proportional to the sine of the solar-aspect angle. Sky coverage is delivered by slow precession of the spin axis around the Sun. This generates observations in a series of loops around the Sun. These loops must (just) overlap, with the consequent large angle of intersection ensuring every point on the sky is observed at large differences in position angle from loop to loop, allowing precise 2-dimensional coordinates to be derived from the essentially one-dimensional measurements. Loop-overlap ensures that Gaia observes any object in the sky at least six times per year, requiring 5.8 revolutions per year, for a precession period of 63days, and precession rate 4deg/day relative to the stars. The precession generates across-scan image motion varying sinusoidally with nominal period 6 hours, and amplitude 173 mas/sec. An allocated window on the CCD containing an image can move as much as 4.5 pixels in the across scan direction during a single CCD transit.

This has an important implication for analysis of Gaia data searching for short-timescale flux variation. The amount of across scan drift depends on the across scan position of the source and the scanning angle, and is not the same for sources at the same location in the focal plane but observed in different telescopes. Windows may suddenly enter in conflict with windows from the other telescope. If a window of a source in one telescope field of view crosses over a window from a source from the other telescope, this can create apparent brightness variations at the CCD level dubbed parasitic sources. The effect depends on the relative positions of the electronic windows on the CCDs, which are unlikely to be exactly the same for different transits: due to the scanning law Gaia employs and because windows are assigned without memory, the window position for a given source can be different for different transits. This can cause an apparent transient brightening in the lightcurve. An example of such an event is shown in Figure 6.

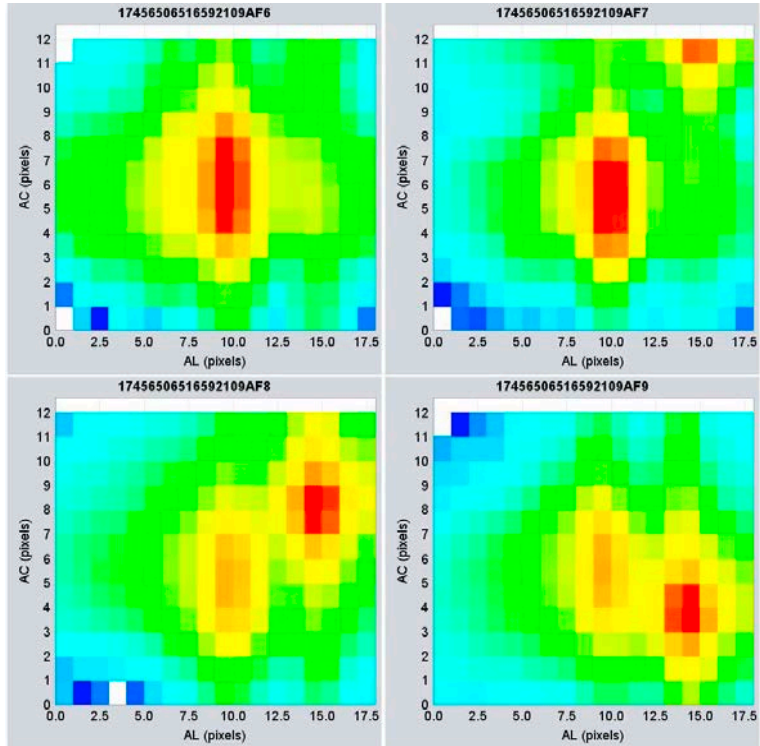

Figure 6. Top left: Gaia image of a relatively bright single star. This illustrates the system point spread function. Top right and lower: a "parasitic" image: a star being observed by the other telescope moves across the focal plan disturbing the flux and position measurement of the original source. Image following [5].

### 2.3 Spacecraft operational issues

The overall status and performance of the Gaia spacecraft and its very many subsystems since launch has remained near nominal expectation until the time of writing (late 2017), auguring well for a successful mission extension beyond the nominal 5-years of operation. However a small number of high profile performance issues became apparent during the (6-month period of) in-orbit commissioning after launch and are noted here.

2.3.1 *Telescope throughput.*

Moist air escaping from the multilayer insulation blankets and/or the carbon-fibre reinforced polymer structural components led to contamination of the mirrors by water ice. This primarily impacted overall system throughput, but also the image quality and the spectral response of the Gaia "G" passband. Throughput losses as high as 30percent were recorded. The ice was sublimated by heating the mirrors, with subsequent several-day intervals required to recover thermal equilibrium, while image quality was recovered with re-focussing. By late 2017 the outgassing was no longer an issue. Considerable effort in photometric and spectrophotometric calibration was required to manage the early rapidly-changing position-dependent sensitivity variations, but has been successfully implemented, so this issue should not affect science users except for the changing passband, which will affect sensitive photometric analyses of time-domain data.

2.3.2 *Scattered light*

Straylight levels seen at the Gaia focal plane are some two orders of magnitude higher than anticipated, and highly variable. The dominant cause was quickly recognised to be sunlight scattered around the edge of the sunshield by bundles of reinforcing fibres which protrude from the sunshield at places where the sunshield edge is not taped, due to movement requirements during deployment after launch. Additional sources are integrated skybrightness reaching the focal plane through unbaffled paths, including a direct illumination contribution through a narrow cone which on occasion scans across bright stars or planets. Mitigation in part has been implemented by careful re-optimisation of the focal plane electronic windows around faint sources, and the source detection algorithm, especially for the radial velocity spectrometer instrument. While adding complexity to data processing and calibration, the main effects of these scattered light sources are to add noise to the faint sources. Fortunately the potentially lost science is mitigated in large part by the increase in dynamic range and sensitivity relative to baseline which is being delivered by Gaia, and by an increase in mission duration beyond the 5-year nominal planned mission – see below.

### 2.3.3 Basic angle variations

Oscillations of the basic angle between the two telescopes were designed to be suppressed at the microarcsecond level, thus providing a robust zero point for absolute astrometry. Nonetheless, periodic oscillations are seen at the 1000microarcsec (preceding telescope)/ 200microarcsec (following telescope) level, with period in phase with the spacecraft 6-hour rotation. An additional 24-hour modulation is also seen, as is modulation correlated with increased thermal activity of the spacecraft video processing units during observations of dense fields, and some other discontinuous effects. The dominant cause is clearly thermo-elastic coupling between the sun-side service module and the shielded payload module. The detailed energy transfer path remains to be fully understood. Fortunately these variations can be calibrated out during astrometric processing, so that the success of the Gaia mission is not imperilled.

There are in addition to these oscillations frequent small jumps, termed "micro-clanks". Their signature is a rapid change in spacecraft rotation rate by up to a few microarcsec/sec, with a rapid return to the pre-excursion rate. Excursions of this type were first identified by van Leeuwen in HIPPARCOS data, and are ascribed to minute structural adjustments (mass displacements) in the spacecraft structure affecting its orientation. They differ from micro-meteoroid (dust grain) impacts which change spacecraft angular momentum, and require micro-thruster correction. Micro-clanks are readily identified and their effects are corrected during data processing.

### 2.3.4 Radiation environment

Radiation damage to electronics and CCD detectors is an inherent aspect of satellite operation. The most significant for Gaia is displacement damage arising from heavy particles – neutrons and protons – causing defects in the crystal lattice with buildup of bulk charge traps. The radiation depends sensitively on the Solar activity level. At times, such as now, of moderate to low Solar activity the dominant source of damage is the slow accumulation of hits by extra-solar system cosmic rays. These are complemented by occasional major Solar flares (CME-events) which can cause as much damage in a single event as many months of steady cosmic ray activity. Very extensive dedicated tests to determine the actual performance and damage level of Gaia's CCDs and electronics are a routine aspect of performance monitoring. To date the relatively low Solar activity level compared to that seen in the last Solar cycle, and used for performance prediction, means that Gaia's CCDs are operating with a very much lower level of radiation damage than pre-launch predictions. Radiation damage, and its corresponding charge transfer inefficiency is not anticipated to be a lifetime limiting factor for Gaia.

### 2.3.4 *Limits on Gaia's operations*

The Gaia spacecraft and all of its critical subsystems continue to perform very well. Gaia's useful data taking lifetime remains limited by design and operational funding availability rather than external accident or system failure. One basic limitation is the supply of (cold nitrogen) fuel for the micro-newton level thrusters for fine attitude control: these maintain Gaia's stable spin rate and orientation at the required milli-arcsecond per second and milli-arcsecond levels respectively. On current estimates, assuming no major changes in system performance, the cold fuel should last until late 2024, by which time Gaia would have achieved a 10-year data taking life. The other limitation is funding to maintain the mission and its (ground) data processing. Maintaining that support requires continual support from the scientific beneficiaries, the community.

### 2.4 *data processing*

Gaia observes on average 1million sources per hour, obtaining all of detection (SkyMapper), white-light photometry and astrometry, blue- and red-spectrophotometry, and for a brighter subset R=11700 spectroscopy near 850nm. The data volume generated by Gaia is significant, with updated information available at https://www.cosmos.esa.int/web/gaia/mission-numbers. During 2018 the number of astrometric CCD measurements will pass one trillion. Each measurement is the crossing time of a target crossing the focal plane, and its associated electronic charge. The observation times represent the one-dimensional along-scan stellar positions relative to the spacecraft instrument axes. The astrometric positions are built up from a suitably large number of such observation times covering a mission observation lifetime long enough to determine the basic parameters of object parallax and proper motion (Fig. 2), and a wide range of scan angles, through an astrometric global iterative solution. This process involves simultaneous reconstruction of the instrument's orientation as a function of time, with a time resolution of two micro-seconds, and the (time-dependent) geometric calibration mapping the detector pixels through the telescopes onto the celestial sphere. It is the feature of this process that the (nuisance) parameters which describe the spacecraft attitude and geometric calibration are derived simultaneously with the astrometric parameters of the astrophysical sources which make Gaia a self-calibrating mission. The very large numbers of observations makes this both technically challenging and mathematically viable, with many hundreds of observations per degree of freedom.

After arrival from the spacecraft Gaia data receive immediate daily pre-processing, followed by cyclic iterative processing to convergence. The daily processing monitors the spacecraft health, performs a preliminary astrometric analysis to assign each source its previous identification if previously observed, and prepares the data for distribution to the data processing centres. It is this early treatment which facilitates the Gaia science alerts discussed below.

The Gaia data processing dataflow follows this scheme.

1. Raw data, reconstructed from satellite telemetry, are analysed utilising a (colour-dependent) modelled point spread function, correcting for CCD detector "features" (charge smearing, electronic noise and zero offset) and astrophysical background flux, to generate a position and flux for each source.
2. These data generate a pre-processing source list.
3. From this, image fluxes (including spectrophotometric data) are refined in Photometric Processing, followed by extensive self-bootstrapped calibration, to provide calibrated source flux and colour.
4. Image location data, and colour data, are used in Astrometric Processing to derive a geometric instrument calibration, and a spacecraft attitude model. The basic five astrometric parameters are derived in this process for every source, including also radial velocity measures (from Gaia's spectrometer) for sources with significant perspective acceleration.
5. The Radial Velocity Spectrometer spectra are reduced and analysed to deliver radial velocities.
6. The calibrated photometric and astrometric data are iterated back to improve step one above.
7. The process is iterated to convergence.

Special efforts are required to manage very bright stars, which saturate the detectors; very crowded regions of sky, which saturate the on-board processing capability; overlapping sources, which are often physically associated binary/multiple stellar systems; resolved sources, typically galaxies or high surface brightness nebulosity; moving objects, typically asteroids, and transient sources, are treated specially.

This data processing scheme involves specialist groups and data centres across Europe, organised into nine Coordination Units supported by six data processing centres, which are located at ESAC, near Madrid (which also hosts the main database), Barcelona, Cambridge, Geneva, Torino and Toulouse.

## 2.5 transient and rapidly moving sources

Gaia repeatedly scans the sky, detecting and measuring all sources that are bright enough. Astrophysical sources can rapidly change substantially in brightness and become "visible" for Gaia. Those transient and flux-variable sources are the subject of a special data analysis, the Gaia photometric science alerts, operated at the Cambridge Institute of Astronomy Gaia Data Processing Centre. The Gaia Science Alerts system detects these transients and provides preliminary type classifications – for example supernova type, cataclysmic variable, tidal disruption event, gravitational microlens, etc from Gaia data. A typical latency between observation and publication is 1-2days. The detections are published for rapid follow-up on-ground both by professional astronomers and the wider public through a dedicated system at https://gaia.ac.uk/alerts. While large numbers of Gaia alerts have been studied and reported in the specialist transient-science publication system (https://wis-tns.weizmann.ac.il/ ) the first refereed scientific article based on Gaia data was a followup of Gaia alert Gaia14aae, an extremely rare example of a totally-eclipsing AMCVn cataclysmic variable[3]. A special feature of this study is that Gaia's discovery was followed up by amateur astronomers, who discovered the unusual nature of the source, motivating further studies, and sharing authorship[4] of the refereed publication [6].

A second type of special source is asteroids, with new examples typically detectable by their rapid motion over the Gaia focal plane, and very many, especially main-belt asteroids, observed specifically based on predicted positions from orbit calculations. New asteroids, and those of special scientific interest, require rapid supplementary ground observations to determine their orbits. As with temporal transients these sources are also published openly and rapidly to allow dedicated (professional and amateur) follow-up, through http://www.gaiagosa.eu/. An interesting feature of Gaia's asteroid detection programme is that Gaia, being beyond the Earth, observes interior to the Earth's orbit, into the orbit of Venus. Thus Gaia has the (unique) capability to discover Earth-crossing asteroids, coming out from the Sun-side (daytime) of the Earth's orbit, which are those potentially providing the greatest threat of collision with Earth.

### 3. Gaia data and data releases

The scientific interest in Gaia includes almost every branch of astronomy, from solar system studies of asteroid mineralogy, through stellar evolution, galaxy evolution, and cosmology to fundamental physics. The Gaia astrometric reference frame, critical to allow source cross-identification from study at other wavelengths or non-electromagnetic processes, also provides critical support to a range of other missions, from control and guidance systems on large ground-based telescopes to fine-navigation for Solar System in-situ missions. The Gaia data are also rich, complex, diverse and are unfamiliar to many in the astronomical community. For all these reasons it was decided very early in the mission studies that Gaia data should be released with no restrictions on use by nationality. Additionally, the science yield is so broad and diverse that it is unrealistic to imagine that a specific group would have the resources to explore the full potential of the whole data set, or even to quantify the limitations of the data quality.

---

[3] An AMCVn system is a short-period binary system in which a hot white dwarf star accretes hydrogen-poor material from a nearby companion. The star AM Canum Venaticorum is the prototype.
[4] https://academic.oup.com/mnras/article/452/1/1060/1748089/Total-eclipse-of-the-heart-the-AM-CVn-Gaia14aae.

Thus mission approval came with acceptance there would be no "protected" science for internal members of the Gaia data processing team, no proprietary period for personal/group science exploitation, and no restriction to the ESA community.

Some science verification prior to data publication is of course an essential aspect of the quality control of the data processing algorithms. Sanity checking is done, and published in the small number of papers published with each data release, which illustrate the properties and quality of the data, and how they may be analysed.

In addition to these validation checks, which are the remit of Coordination Unit CU9, which manages the data publication, two other Coordination Units focus on derivation of statistical properties of Gaia sources. CU7 analyses the time-series photometry ("light curves") of detected variable stars, providing a uniformly-derived set of basic astrophysical parameters. The primary motivation and global benefit of these analyses is to quality-control the photometric calibration work. By providing cross-checks of special classes of variable stars, especially those used to calibrate the distance scales in astronomy, this work additionally provides invaluable consistency checks on the astrometric reductions, ensuring there are no unrecognised systematics perturbing the data. CU8 models the combined photometric and astrometric stellar data to deduce the basic physical properties of well-behaved stars, providing estimates of stellar effective temperature, line-of-sight extinction, and other physical properties. The radial velocity spectra for sufficiently bright stars are also analysed to provide several elemental abundances. The primary motivation and global benefit of these analyses is to quality-control the combination of the spectro-photometric and parallax/distance-scale calibration work. It is anticipated that the community will develop and implement their own optimised analysis efforts.

Interestingly, and deliberately, the Gaia data releases, which will in future include parallaxes, proper motions, spectro-photometry, radial velocities and astrophysical parameters, do not include derived distances. The reason for this is that the Gaia project is collaborating with the community to develop awareness of the methodologies appropriate to derive robust science from massive astrometric data sets, an unfamiliar tool since previously unavailable. Specific example science analysis articles are published with the data releases as a fundamental part of this information transfer, specifically to introduce the importance of using the error correlations appropriately. An example tutorial article is [7]. This makes the point that experiments measure parallax, with uncertainties, while distance is the inverse of parallax. Inversion of a parallax is almost never the optimal way to deduce distances, and can fail numerically since noisy parallaxes can be negative, while distances cannot. Rather deduction of distances is an inference problem, in which assumptions and priors are unavoidable. Similar strictures apply to conversion of parallaxes and proper motions to space motions.

After the first full five-years of mission operation the first full release of calibrated Gaia data will be made public. This will include every Gaia observation, allowing for the first time the community to handle the full information content of Gaia data optimally, by applying their own astrophysical models. There are many obvious examples of key scientific studies which will be feasible only by re-modelling the Gaia individual measures. As one example, many, possibly all, stars are multiple, in having physically associated companion stars and/or planetary systems. The default Gaia model however fits the standard five astrometric parameters, which assume all stars are isolated single sources. This will absorb long-period orbital motion in the proper motion, and will capture low-amplitude short period orbital motions in (marginally) significant residuals. A billion sources generate a lot of false positives, so finding the signal in the noise is not trivial. Generally, if the 5-parameter fit is statistically adequate no multiple-source model is attempted in the Gaia processing system. One can manifestly do much better, and will need to, to find, for example, the several tens of thousands of local exo-planetary systems by jointly analysing the full Gaia astrometry and photometry data sets. This data set will however not be calibrated and able to be released until, on present plans, 2022/2023. Cornucopia awaits. Manifestly, handling, analysing and visualising multi-dimensional data with complex correlations for what will eventually be over 2 billion sources is not trivial. That subject merits an article on its own.

### 4. Gaia science – an overview

The scientific case to support the Gaia mission approval was developed with substantial community support during the 1990s, and published as the ESA Concept and Technology Study Report "Red Book" [8] which can be accessed through the ESA Gaia web site. A summary was published in 2001 [9]. The summary of those articles is of interest, in showing that the scientific and technical aims of Gaia remain state of the art, and that none of the science goals which depend on space astrometry has been significantly advanced in the interim from ground or other mission results. The aspect which has changed most is interest in photometric variation in astrophysical sources on all observable time-scales (time-domain astronomy), which has over the last two decades become a very major activity in its own right. The summary of the Red Book is worth repeating here: *"GAIA[5] will provide positional and radial velocity measurements with the accuracies needed to produce a stereoscopic and kinematic census of about one billion stars throughout our Galaxy and into the Local Group, amounting to about one per cent of the Galactic stellar population. GAIA's main scientific goal is to clarify the origin and history of our Galaxy from a quantitative census of the stellar populations. It will advance questions such as when the stars in our Galaxy formed, when and how it was assembled, and its distribution of dark matter. The survey aims for completeness to V=20mag with accuracies of about 10µas at 15mag. Combined with astrophysical information for each star provided by on board multi-colour photometry and limited spectroscopy, these data will have the precision necessary to quantify the early formation and subsequent dynamical, chemical and star formation evolution of our Galaxy. Additional products include detection and orbital classification of tens of thousands of extra Solar planetary systems, and a comprehensive survey of some $10^{5-6}$ minor bodies in our Solar System, through galaxies in the nearby Universe, to some 500,000 distant quasars. It will provide a number of stringent new tests of general relativity and cosmology."*

These ambitions remain the current Gaia challenge, with significant progress already reported towards most aspects of the science case, and many

---

[5] The acronym GAIA was still in use at that time.

performance estimates being shown to be conservative.

Since the first release of very preliminary Gaia data (GDR1, https://www.cosmos.esa.int/web/gaia/dr1) on 14-September-2016 (1000days after Gaia launch, including roughly one year of observations and 1.1billion sources, but only 5million parallaxes) about one science analysis paper per day has appeared, covering a very broad scientific range.

Rather than attempt a full review of all Gaia science, we note here a few highlights which are anticipated to show rapid advance following the next data release (GDR2, 04/2018) and later.

### 4.1 The distance scale.

Calibrating cosmological distances remains a state of the art challenge in astrophysics, with both the consistency of the standard ΛCDM cosmology model, and the determination of an equation of state for Dark Energy being key challenges. This is typically presented as precision determination of the expansion rate of the Universe, the Hubble constant, and its variation with time/distance. At present a marginally significant tension exists (~3.7standard deviations) between the local expansion rate required by the standard model of cosmology normalized by observations of the cosmic microwave background and that determined directly. Gaia's contribution here will be substantial, essentially by allowing precise fundamental calibration of all the many primary distance calibrators, thus allowing the first robust analyses of their systematic and random dispersions. Most of these calibrators are essentially stars in short-lived or terminal evolutionary states. These stars are intrinsically luminous, so can be seen at large distances. They also can have some property which is related to intrinsic luminosity which can be measured with affordable observational effort. The most important are pulsating variables, where the pulsation period can be determined from simple repeated flux measures, and the intrinsic luminosity deduced.

Stars which are pulsating variables (Cepheids, which are high-mass young stars, RRLyrae, which are old low-mass stellar cores) have a calibrate-able relation between pulsation period and intrinsic luminosity (Leavitt's Law[6]) allowing their use as Standard Candles. Similarly, although more complex, other pulsating variable stages late in a star's life (Asymptotic Giant Branch, Sun-like stars near the end of their life) are also valuable, as is the limiting luminosity of the Red Giant Branch evolutionary phase for low-mass stars (RGB-tip). These tracers include young (Cepheid), old (RRLyrae), and intermediate age (RGB-tip, AGB) evolutionary phases and so provide complementary consistency checks. All when combined calibrate in turn Type Ia supernovae in nearby galaxies in which several different calibrators can be observed at the same distance. Type Ia supernovae are sufficiently intrinsically luminous to extend distance calibration to sufficiently large distances to probe the entire range over which Dark Energy has been the dominant energy-balance term in the Universe. At present luminosity/distance calibration of the standard candle tracers is the largest contributor to the error budget. Post-Gaia that contribution will be essentially removed, leaving corrections for interstellar extinction and the diversity of Ia supernovae as the dominant contributors.

---

[6] The relationship that Cepheid variable stars have a pulsation period which is correlated with their intrinsic luminosity, longer periods being more luminous. This was discovered by Henrietta Leavitt working at Harvard College Observatory in 1912.

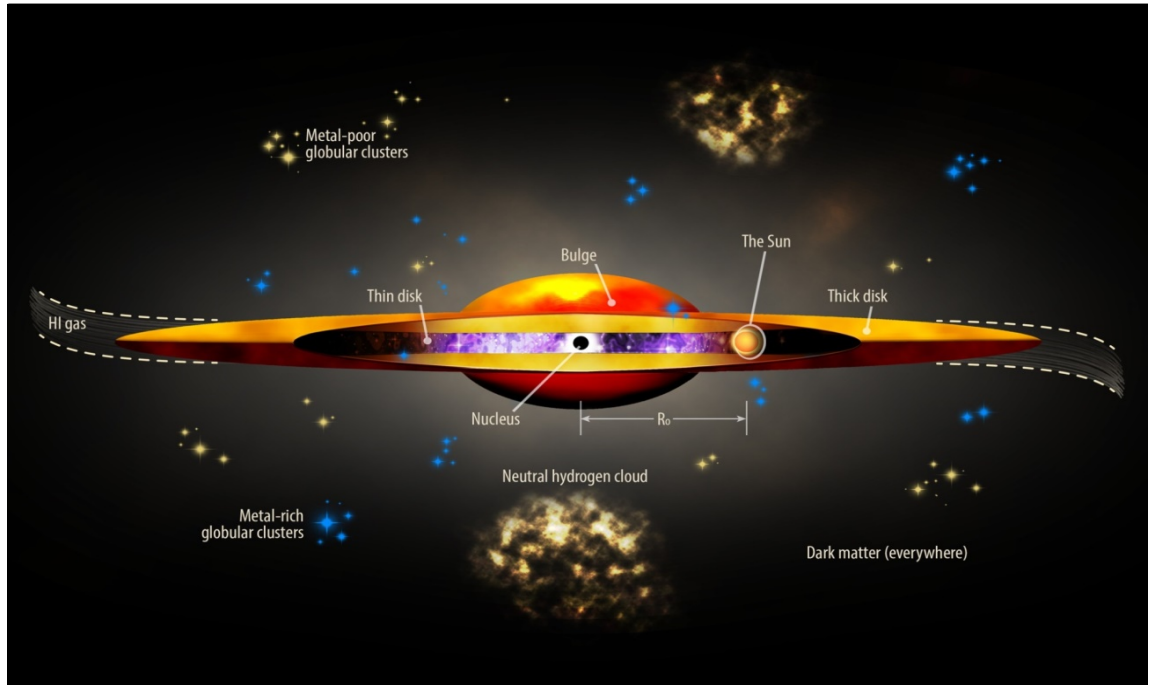

Figure 7. A cartoon Milky Way, with the primary structural components discussed in the text being identified.

4.2 Milky Way structure and evolution

At a local galaxy level distance calibrations also provide key astrophysics, especially in quantifying the properties of the far outskirts of our Galaxy, which are dominated by debris from minor galaxy accretion and disruption across time. The timescale for a small galaxy or star cluster to be tidally unbound and have its member stars mixed into the field is comparable to the age of the Universe in the far outer Milky Way. Thus every merger or accretion/disruption event remains visible in phase-space as an overdensity, or in configuration space as a stream. The accretion/assembly history of the Milky Way can be reconstructed by counting how many distant stars are in streams once their distances, space motions and chemical abundances are knowable. The outer Galaxy is indeed very lumpy, quantifying the accretion rate over time.
Timescales are shorter in the inner Galaxy, so that although recent mergers can be seen directly, more indirect methods must be applied for earlier events.

Most effective is to complement phase-space studies with detailed chemical elemental analyses (Fig 1.). In essence, the abundance of elements created in stars (everything heavier than Helium, generically misleadingly termed "metals" in astrophysics) increases with time in a closed evolving system, but can change discontinuously in a merger event. One therefore maps the dependence of sensitive element ratios as a function of some element generated only in common (Type Ia) supernovae. The ratio of Mg to Fe as a function of the ratio of Fe to H is typical. Scatter/discontinuity implies mergers, smooth correlations imply star formation and self-enrichment in stable gas. Combining this information with spatial and phase-space distribution functions is the basic analysis methodology for stellar archaeology. From such studies we have learned that the 3-component view of the Galaxy – disk, bulge, halo – is more complex (Fig 7). Indeed all three stellar population components have at least two contributions. The galactic disk is made up of (at least) a young thin disk and an older thick disk, with local stellar dynamics also showing evidence of a massive central bar in the disk. The inner bulge contains both a very old population, and a dynamically-generated X-shape structure of inner disk stars. The halo contains (at least) both a substantial old population apparently formed *in situ*, and the later accreted population. Quantifying the relative contributions and properties of these many sub-systems is a primary role for Gaia data analyses. It is already clear that the simplest prediction of standard cosmology, that the outer parts of our Galaxy are dominated by the merger and accretion of small sub-systems, is supported by observation. However, it is equally clear that the bulk of the stars in the Milky Way have formed in situ over many Gyr from gas which has been systematically enriched in chemical elements from stellar evolution and supernovae in the Galaxy. This implies that baryons have accumulated into the Milky Way as gas, and formed stars only later. The merger history of the Milky Way has been very quiescent over most of its history. Reproducing this history remains a challenge for galaxy evolution simulations.

On larger scales Local Group galaxies are having their 3-D orbits quantified, providing valuable tests of the kinematic evolution of a now-bound system since very early in the Universe. Indeed the outer Local Group is the closest place where we can observe space-time dominated by expansion driven by Dark Energy, whatever that is. Precision measures of dynamics in that regime will be one of few available probes of the equation of state of the Dark Energy.

One of the primary motivations for Gaia is to provide a high spatial resolution map of the distribution of mass, including the Dark Matter, whatever that is. Through application of the coupled Poisson-Boltzmann equations, combining 3-D spatial density distribution and 3-D kinematics, it is feasible to deduce pressure scale lengths, and quantify the balance between pressure generated by stellar kinematics and the gravitational potential gradients which balance them. This is determination of mass distributions, and discovery and mapping of cold dark matter. Does dark matter have any associated physical scale, which would indicate some physical property? Essentially, a minimum speed would correspond to a temperature, an associated characteristic spatial scale length, and a particle mass-velocity-density relation, clues to its/their nature. Among the questions we hope to answer are the existence, or not, of dissipational dark matter – which would form a very thin disk distribution in the Plane of the Galaxy. Current indications do not favour this. Similarly, vanilla CDM assumes no physical properties associated with CDM particles, so they will form gravitationally bound systems on scales as small as a lunar mass. Very large numbers of low-mass bound dark matter systems are predicted in this case. Can we detect them, by their tides, wakes, damage to low-velocity stellar systems, or otherwise? This is the really big challenge for the next few years, whose results we await with interest.

### 4.3 Stars and their end-points

The most straightforward product of an astrometric survey is determination and presentation of the true intrinsic luminosities and colours of stars. These in astronomy are conventionally displayed as an Hertzprung-Russell diagram[7] (Fig. 8). The information in an observational Hertzprung-Russell diagram is that needed to quantify stellar evolution, a subject which is in consequence rapidly increasing in interest. Among questions which can be addressed by the very large sample in Gaia are short-lived evolutionary phases, particularly those involving very young stars, those involving high-mass (necessarily also young) stars, and rapid and poorly-understod rapid evolutionary phases as stars transition between relatively stable states. Quantification of the time-dependent evolutionary phases involving two high-mass stars which transfer mass back and forth as they evolve is essential to understand the origins of those multi-stellar mass black hole binaries recently observed as gravitational wave sources.

Stars form in molecular clouds, turning into clusters which evolve dynamically in the Galactic tidal field, dissolve into streams and moving groups, and eventually form the field star populations of the Milky Way. All these dynamical processes are complex, and require high quality high-precision luminosity and kinematic information for their quantification. These are the primary outputs from Gaia.

---

[7] The Hertzprung-Russell diagram was derived in modern form in significant part by Henry Norris Russell when a fellow at Cambridge (UK). His paper [10] completed a project to determine stellar parallaxes using the new technology of photography. This had been initiated by the local astronomer Sir Robert Ball in 1893, and utilised a special-purpose telescope funded from an endowment established by Anne Sheepshanks [11].

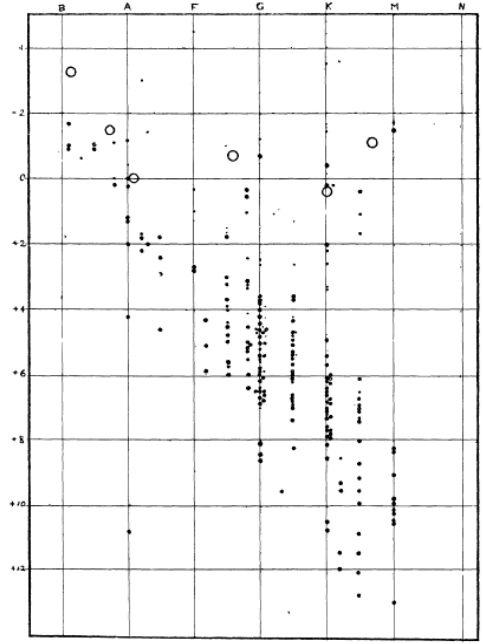
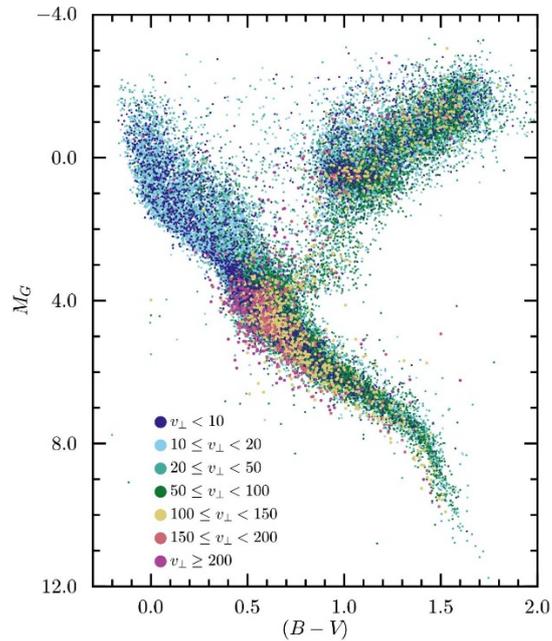

Figure 8 – LHS. An early presentation of the Hertzprung-Russell diagram by Russell. RHS the comparable diagram from Gaia's first data set, with stars colour-coded by transverse velocity (from Gaia Collaboration 2016 [12]).

4.4 Gaia's spectrophotometry

A very powerful information set from Gaia which has the potential to be as revolutionary as will be the parallaxes is the Gaia spectrophotometry. This is obtained through two prisms, with the blue system having dispersion ranging from 3 to 27 nm/pixel over the range 330-680nm, and the red system ranging from 7 to 15 nm/pixel over the range 640-1050nm. The resulting time-series spectrophotometry of every object observed by Gaia is a completely new concept in astrophysics, whose impact will be very substantial. As just one example of this dataset figure 9 shows the spectrophotometric times series for a supernova, Gaia16aeg. The figure shows the developing supernova spectrum, from hot plasma until visibility of the lines characterising the newly created chemical elements being distributed into the interstellar medium of the host galaxy. The corresponding data for Solar System asteroids classifies every asteroid into its appropriate mineralogical family, and more generally allows identification of the astrophysical type of Gaia sources, from stars to quasars.

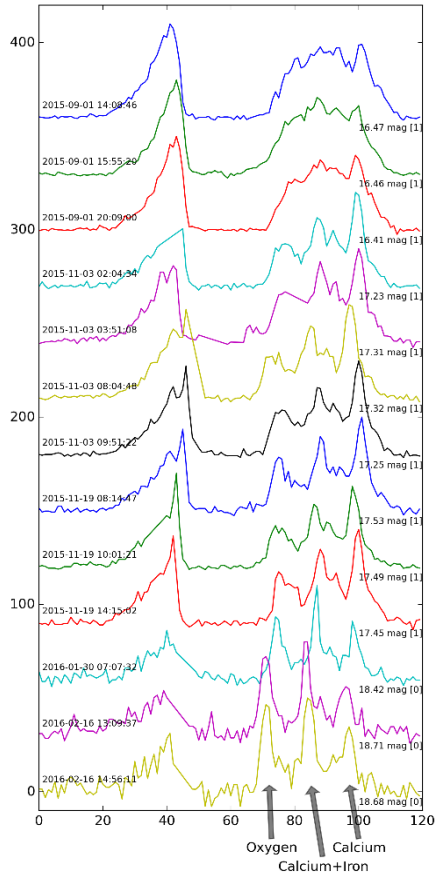

Figure 9. This sequence of spectra charts the evolution of Gaia16aeg (ASASSN-15lv) as it transitions towards a nebular spectrum. Over a five month period, the continuum fades while strong emission lines of Calcium (including the Ca near-infrared triplet), Oxygen and Iron are seen to emerge in the red spectrum. Courtesy ESA.

4.4 **Fundamental physics**

Relativistic effects are highly significant for Gaia measurement accuracy, with tests of General Relativity being a significant driver from the very start of the project. This established tight constraints on the mission. For example, sufficient modelling of Newtonian aberration requires that the spacecraft orbit (Lissajous orbit around L2) is quantified with a velocity accuracy of 1 mm/s. Finite light velocity effects lead to position-dependent propagation delays in the field of view which must be accounted for. Monopole light deflection (the famous 1.75arcsec solar limb effect first verified by Eddington & Dyson in 1919) exceed the microarcsec level all-sky for the Sun, and up to 90deg from Jupiter, significantly complicating the computational effort. Quadrupole light bending is 240µas at the Jupiter limb, and is 1µas at 8 Jupiter radii. This allows a special Gaia experiment – to quantify light bending by Jupiter, this test involving an oblate rotating mass moving in a deeper (Solar) potential. The much improved accuracy of asteroidal orbits from Gaia astrometry is another significant improvement in a classical GR test. The wide range of relativistic tests feasible from Gaia data, and their implications for Gaia data processing, are introduced in [13] and [14].

One potential Gaia sensitivity which has become very topical is detection of gravitational waves. Solar mass binary black holes during their final merger phase have been detected with ground based systems (LIGO, VIRGO). However supermassive black holes, which are known to exist in essentially all galactic nuclei, in binary merger evolutionary phases which must follow all galactic mergers, radiate at lower frequencies, inaccessible to ground-based instruments. The planned space mission LISA will detect merging black hole binaries in the mass range 10^5 to 10^7 solar masses out to very high redshifts. More massive binaries in their early inspiral stage can potentially be detected by pulsar timing arrays and by Gaia astrometry. Fig 10 illustrates the signal, and is from [15] who discuss the options further. See also [16].

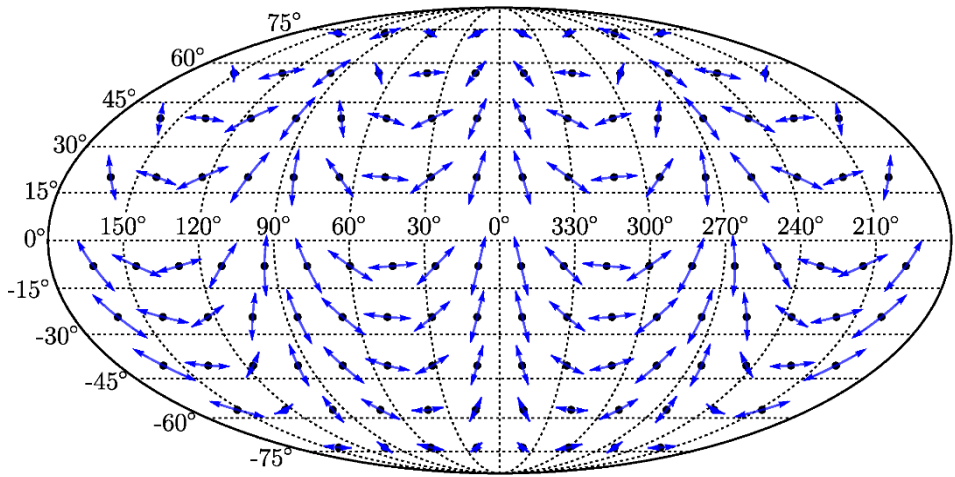

Figure 10. The signature of a constant frequency gravitational wave propagating from the direction of the north galactic pole. For clarity the gravitational wave has an unphysically large strain amplitude. The fourfold rotational symmetry of the transverse-traceless gravitational wave is evident. Figure courtesy of [15].

5. **Conclusions**

The ESA Gaia mission is opening one of the last remaining observational windows to the Universe. Gaia is the first mission measuring large numbers of parallaxes to faint magnitudes, with well over 2 billion sources being observed to magnitudes sufficiently faint to link distance scales in our Solar System directly to cosmological scales. These parallaxes may be modelled to deduce their inverse, distances. Gaia additionally provides high-precision proper motions for every observed source, allowing derivation of transverse speeds. For the 300million brightest stars, Gaia additionally provides a line-of-sight radial velocity, so that for those stars 3-D space velocities are derivable. In addition to this astrometry Gaia provides very precise photometry, with typically 16 observations per year (up to many times this in some areas) each year over the up to10year mission life. This allows detection and modelling of variability, and detection of transients and rapidly moving Solar System objects. There is even more: Gaia is delivering the first ever spectrophotometric survey, with every object having an energy distribution determined at every observation.

Information of this type, accuracy, precision and on this scale has never before been available. In addition to its remarkable strong and broad scientific case building our understanding of the Universe, one may be confident that new discoveries and new questions will follow.

Acknowledgement.
This article has made use of results from the European Space Agency (ESA) space mission Gaia, the data from which were processed by the Gaia Data Processing and Analysis Consortium (DPAC). Funding for the DPAC has been provided by national institutions, in particular the institutions participating in the Gaia Multilateral Agreement. The Gaia mission website is https://www.cosmos.esa.int/web/gaia/home. Very many people continue to work to ensure Gaia data are obtained, reduced, calibrated, and published for open and free use by the global community. We are all in debt to them all (Fig 11). The author acknowledges partial support from European Research Council grant 320360, and Dr. Gudrun Pebody for references to Hesiod and help with Figure 11.

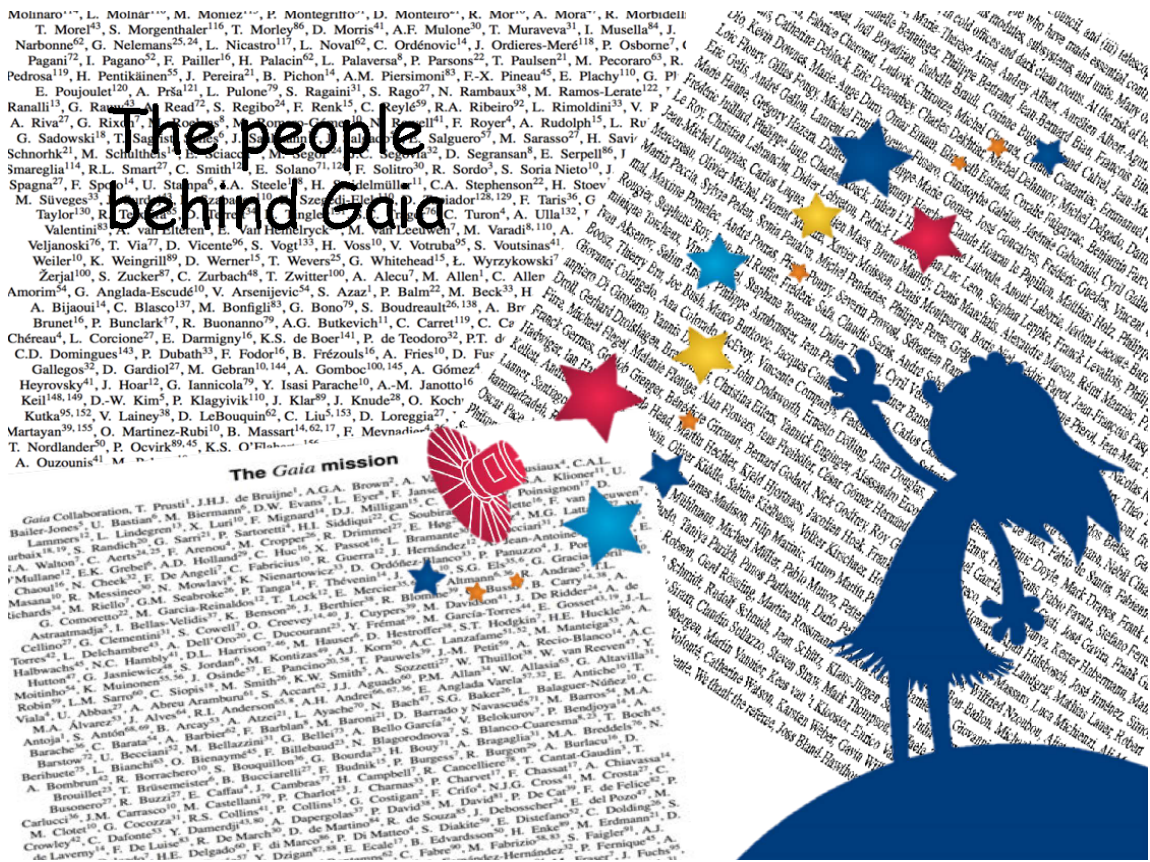

Figure 11. A mission on the scale of Gaia has involved dedicated work already for 25years from many hundreds of people. These include the spacecraft design, build and test teams, and the scientific specification and data processing and analysis teams. An incomplete list of some 1000 of the people who make the Gaia mission a success is shown here.

Acronyms and Glossary

AF   Astrometric field – the main part of the focal plane illustrated in Fig. 5.
AGB   Asymptotic Giant Branch.  A short-lived luminous cool evolutionary state occupied by all low to intermediate mass stars (less than about 10 Solar masses) late in their lives.
AGIS   Astrometric Global Iterative Solution. Calculation of basic 5 astrometric parameters for every star and the full description of the satellite orientation and optical model from Gaia observations.
AM CVn   An AMCVn system is a short-period binary system in which a hot white dwarf star accretes hydrogen-poor material from a nearby companion. The star AM Canum Venaticorum is the prototype.  cf footnote 3.
Basic angle   The angle (106.5deg) between the primary mirrors of the two Gaia telescopes. cf Fig 4.
BAM   A laser interferometer which measures the basic angle
BP          Blue (Spectro)Photometer. A low-dispersion prism plus filter providing slitless spectrophotometry with variable resolution between 330nm and 680nm.
CCD   Charge Coupled Device.  An integrated circuit etched onto a silicon surface forming light sensitive elements called pixels. Photons incident on this surface generate charge that can be read electronically.
CDM   Cold Dark Matter. An unknown form of gravitating matter which dominates the mass budget of the Universe. As yet undetected directly, it is not inconsistent with being varieties of very weakly interacting fundamental particles.
Cepheid   An evolved high-mass star in an evolutionary state such that it pulsates radially, varying in both diameter and temperature and producing changes in brightness with a well-defined stable period and amplitude.  cf Leavitt's Law.
CME   Coronal Mass Ejection.  A significant release of plasma and magnetic field from the solar corona into the solar wind. CMEs generate aurorae, and damage spacecraft electronics.
CU   Coordination Unit – a short name for a team with specific responsibilities inside the Gaia Data Processing and Analysis Consortium.
Dark Energy   An unknown form of energy which dominates the energy budget of the Universe. Its effect is to accelerate space-time against gravity.
DPAC   Gaia Data Processing and Analysis Consortium. The teams who process and publish Gaia's data for community analysis. Say thank you.

ESA    European Space Agency

ESAC   European Space Astronomy Centre is ESA's centre for space science. It hosts the science operation centres for all ESA astronomy and planetary missions together with their scientific archives.  It is located near Madrid.

G  short for G-photometric passband, the natural white-light sensitivity passband of the Gaia astrometric CCDs.

HIPPARCOS      an acronym for HIgh Precision PARallax COllecting Satellite, an ESA mission which proved the viability of space astrometry, and acted as a pathfinder for Gaia. Appropriately the pronunciation is also very close to Hipparchos (Ἵππαρχος) the name of a Greek astronomer who lived from 190 to 120 BC.

HRD   cf. footnote 6. Hertzprung-Russell diagram, a plot showing intrinsic stellar luminosity (derived from parallax) and intrinsic colour. Fig. 8 is an example.

Leavitt's Law  The relationship that Cepheid variable stars have a pulsation period which is correlated with their intrinsic luminosity, longer periods being more luminous. This was discovered by Henrietta Leavitt working at Harvard College Observatory in 1912.  Cf. footnote 5.

Lightcurve  A plot of apparent brightness vs time for an astronomical source.

$\Lambda$CDM  the standard cosmological model, involving General Relativity, and two components containing 95% of the energy density of the Universe which are not understood. Nonetheless it describes large scale structure very accurately.

Parsec   Parallax defines the basic distance unit in astronomy, the parsec, that being the distance corresponding to a heliocentric parallax of one second of arc, equivalent to 206265 times the astronomical unit, the mean Sun-earth distance, or 3.0867E+16m, or 3.262  light years. Footnote 1.

PSF/LSF  Point Spread Function/Line Spread Function. The 2-D/1-D energy distribution in an image in the focal plane of a telescope.  Figure 6 shows the Gaia PSF.

RGB   Red Giant Branch. The evolutionary stage for a low-mass star when core-hydrogen burning has ended, and shell-hydrogen burning supports the star. The star becomes large, cool, luminous and red.

RP           Red (Spectro)Photometer. A low-dispersion prism plus filter providing slitless spectrophotometry with variable resolution between 640nm and 1050nm.

RR Lyrae   A stellar evolutionary phase following the RGB in which the star is hot, pulsating, and follows a period-luminosity relation.

RVS         Radial Velocity Spectrometer,  a transmission-grating spectrometer which is part of the Gaia payload. It delivers R~11700 spectra in the interval 845nm to 872nm

SIF  Service Interface Function. A Gaia spacecraft operational mode in which images may be obtained, overriding the usual TDI observational mode.

SM/SkyMapper   SkyMapper is a column of CCDs dedicated to source detection and analysis in real time, followed by allocation of an appropriate electronic  window to track a detected source in TDI mode.

TDI   Time Delayed Integration. A CCD control mode in which the accumulated charge is clocked across the CCD at the same rate as a moving image, to continue flux integration.

Type Ia   A type of supernova, probably usually caused by a merger of two white dwarfs. The intrinsic luminosities of Type Ia supernovae are correlated with their other observational properties (duration, etc.) and so may be calibrated as cosmological standard candles.

WFS   WaveFront Sensor. A CCD sensor which measures, via  a Shack-Hartmann system, the optical alignments and focus of the Gaia mirrors.